\documentclass[preprint,nofootinbib,aps,superscriptaddress,floatfix]{revtex4}
\usepackage{epsfig}
\def \beq{\begin{equation}}
\def \eeq{\end{equation}}

\begin{document}

\preprint{NUHEP-TH/06-02}

\title{A Three-Flavor, Lorentz-Violating Solution to the LSND Anomaly?}

\author{Andr\'e de Gouv\^ea}
\affiliation{Northwestern University, Department of Physics 
\& Astronomy, 2145 Sheridan Road, Evanston, IL~60208, USA}

\author{Yuval Grossman}
\affiliation{Department of Physics, 
Technion--Israel Institute of Technology, Technion City, 32000 Haifa,
Israel
\vspace*{1cm}}

\begin{abstract} \vspace*{2mm}
We investigate whether postulating the existence of Lorentz-violating,
CPT-conserving interactions allows three-neutrino solutions to the
LSND anomaly that are also consistent with all other neutrino data. 
We show that Lorentz-violating interactions that couple only to one of
the active neutrinos have the right properties to explain all the
data.  The details of the data make this solution unattractive.  We
find, for example, that a highly non-trivial energy dependence of the
Lorentz-violating interactions is required.
\end{abstract}

\maketitle

\section{Introduction}
\label{sec:intro}

In 1996, the LSND collaboration first presented statistically
significant evidence for $\bar{\nu}_e$ appearance from antimuon decay at
rest \cite{LSND_first}. Further analysis of larger LSND data samples
(which also included data from pion decay in flight) confirmed the
original excess \cite{LSND}. In spite of the fact that this so-called
LSND anomaly is close to completing ten years of existence, there is
no compelling physics explanation for it.  It is very natural to try
to understand the LSND anomaly by postulating that there is a
probability for what was originally a muon antineutrino to be detected
as an electron antineutrino. If this is the explanation for the LSND
anomaly, the associated transition probability need not be very large:
$P_{\mu e}\sim 0.2\%$~\cite{LSND}.

Neutrino oscillations are the leading candidate explanation for the
LSND anomaly, for several reasons. First, neutrino oscillations are
known to occur in nature -- it is all but certain that mass-induced
neutrino oscillations are the solution to the solar and atmospheric
neutrino puzzles \cite{TASI}. Second, the $L$ dependence of the
$\bar{\nu}_{\mu}\to\bar{\nu}_e$ oscillation probability is crucial for
reconciling the LSND anomaly with negative results from the KARMEN-2
experiment \cite{karmen}.

Mass-induced oscillations among the three known ``active'' neutrino
species, however, are known not to provide a viable solution to the
LSND anomaly. The reason is very easy to understand. In terms of
two-flavor $\nu_{\mu}\to \nu_e$ oscillations, the LSND data (combined
with other ``short baseline'' data) require $\Delta m^2\sim 1$~eV$^2$
and $\sin^22\theta_{e\mu}\sim 10^{-3}$. On the other hand, the
well-established atmospheric and solar neutrino oscillations require
$\Delta m^2\sim 10^{-3}$~eV$^2$ and $\Delta m^2\sim 10^{-4}$~eV$^2$,
respectively. With three neutrino masses, it is impossible to obtain
three mass-squared differences that differ by orders of magnitude. For
this reason, the existence of so-called light sterile neutrinos (that
couple to the active neutrinos via mixing) is often postulated when it
comes to addressing the LSND anomaly. It is remarkable, however, that
the addition of one sterile neutrino does not provide a good fit to
all neutrino data \cite{4_analysis,5_analysis}, while the addition of
two (or more) sterile neutrinos seems to provide a better fit
\cite{5_analysis}. The goodness-of-fit of $3+2$ models when one
includes all of the world's neutrino data is, however, currently under
detailed investigation,\footnote{We thank Michel Sorel and Osamu
Yasuda for bringing out this point.} and, furthermore, concordance
cosmology fits to all available ``cosmological data'' virtually rule
out sterile neutrinos masses larger than $m_{\nu}\gtrsim 1$~eV if
these mix significantly with the active neutrinos. This is precisely the case
of sterile neutrinos that help resolve the LSND anomaly \cite{nu_cosmology}.

The fact that mass-induced oscillations do not provide a compelling
solution to the LSND anomaly is best illustrated by the fact that
there are several alternative solutions to the LSND anomaly
\cite{alternatives,good_alternatives,decoherence,other_liv,Auerbach:2005tq}.
While many of them \cite{alternatives} are now ruled out by more
current data or more refined data analysis (see for example,
\cite{4_analysis,strumia,concha}), most of the ones that seem to work
require new, very light degrees of freedom (say, sterile neutrinos or
scalars) {\sl and} extraordinary new physics (say, CPT violation)
\cite{good_alternatives}. Potential exceptions include postulating
that CPT-invariance is violated and that neutrino propagation is not
described by the Schr\"odinger equation \cite{decoherence}, or
imposing Lorentz-violating interactions that include may
direction-dependent effects
\cite{other_liv,other_liv_long,Auerbach:2005tq}.  It is yet to be
properly demonstrated whether either of these two latter possibilities
can indeed accommodate all the neutrino data.

It is interesting to explore whether one could resolve -- at least in
principle -- the LSND anomaly without introducing any new, light
degrees of freedom.  Since new
parameters need to be added to the ones present in the
$\nu$SM,\footnote{The $\nu$SM refers to the standard model of particle
physics, plus the addition of mass parameters for the light neutrinos
\cite{theory_reviews}.} we will postulate that one of the neutrinos
couples to a source of Lorentz Invariance Violation (LIV). By varying
the flavor composition of this neutrino and the energy dependence of
the LIV effect, we are able to probe for solutions that qualitatively
accommodate all the neutrino data. The reason for this particular
choice of new physics is straightforward, and will become clear in
the next sections. It provides a minimal number of free parameters to
fit the LSND data, and also contains a mechanism for allowing the rest
of the neutrino data to be properly interpreted in terms of
``ordinary'' mass-induced neutrino oscillations.

Our intentions are two-fold. First, as already stated above, we wish
to determine whether there is indeed such a solution, and how
constrained it is. We found that when only one superposition of
neutrinos is coupled to the Lorentz violating sector the general
characteristic of the data can be reproduced. That is, the Solar
neutrino and Atmospheric neutrino are describe to a very good
approximation by the $\nu$SM. LSND data, however, is taken care of by
the new LIV interaction.
Second, we will use our LIV solution to estimate how much ``fine-tuning''
is required for a three-flavor solutions to the LSND anomaly. 
Here, our main result is rather negative. We find that in order to
accommodate all neutrino data, the LIV model has to be highly
finely-tuned. Thus, while we do report here on our attempts to find a
LIV solution to all neutrino data, our conclusion is that such a
solution is unlikely to be realized in Nature.

The paper is organized as follows. In
Sec.~\ref{sec:formalism}, we present the mechanism we wish to explore,
and how it modifies neutrino oscillations. In Sec.~\ref{sec:fit}, we
qualitatively fit all neutrino data with the model outlined in
Sec.~\ref{sec:formalism}. In Sec.~\ref{sec:end}, we speculate on
possible origins for LIV and summarize our results.

Our study is qualitatively different compared to other studies of
Lorentz invariance violation and neutrino oscillation experiments
\cite{other_liv,Auerbach:2005tq,other_liv_long}. 
Refs. \cite{other_liv} and \cite{Auerbach:2005tq} concentrate on
observable consequences of LIV for ``short'' baseline neutrino
experiments, including LSND. (Ref. \cite{Auerbach:2005tq} very briefly
comments on other neutrino oscillation experiments.) In
Ref. \cite{other_liv_long}, an attempt is made to fit all neutrino
data. Various different scenarios of Lorentz invariance violation are
discussed, but the fit is done for subsets of the neutrino data at a
time. Moreover, only oscillation with $L/E$, $L$, and $LE$ dependent
were considered. Based on our results, we believe that none of the
examples explored in \cite{other_liv_long} can fit all neutrino data
once a global fit is performed. This conclusion is based on the energy
dependency, regardless of whether neutrinos are assumed to be massive,
whether the LIV effects are directionally dependent, or whether CPT is
conserved. In contrast, our emphasis is on trying to qualitatively accommodate {\sl
all} neutrino data, including those from LSND. Unlike other
studies, our main idea is to couple only one neutrino to the LIV
source.  As a specific model we concentrate on
CPT-conserving, direction-independent effects. To fit all data we find
ourselves forced to explore more exotic energy dependent LIV effects.

\section{Formalism}
\label{sec:formalism}

We postulate that one of the neutrinos couples to a source of
Lorentz-invariance violation (LIV). This assumption is motivated by
minimality but turns out to be very important, as will be discussed
later.

In the case of one neutrino with mass $m$, we postulate that LIV
manifests itself via a modified dispersion relation of the form
\begin{equation}
E\sim |\vec{p}|+\frac{m^2}{2|\vec{p}|}+\frac{f(|\vec{p}|^2)}{2|\vec{p}|}.
\label{disp_relation}
\end{equation}
Eq.~(\ref{disp_relation}) is valid as long as $|\vec{p}|\gg
m,\sqrt{f}$. $f$ is some function of $|\vec{p}|^2$, so that CPT is
conserved.\footnote{More generally, we could have chosen $f$ to also
contain odd powers of $|\vec{p}|$. These, however, violate CPT,
and lead to a different modified dispersion relation for neutrinos and
antineutrinos. We find that this extra complication does not add
significantly to our analysis, and we choose to work only with even
powers of $|\vec{p}|$. This choice is natural in the
sense that it is protected by a symmetry (CPT-invariance).} We assume
that $f$ can be expanded as
\begin{equation}
f=2\sum_{n=1}^{\infty} b_n\left(\frac{|\vec{p}|}{E_0}\right)^{2n},
\end{equation}
where $E_0$ is some convenient constant with dimensions of energy,
while the coefficients $b_n$ have dimensions of mass-squared.

Assuming that all approximations above are valid, neutrino
oscillations are described by the effective Schr\"odinger equation
\begin{equation}
i\frac{\partial}{\partial L}\nu_{\alpha}=H_{\alpha\beta}\nu_{\beta},
\end{equation}
where $\alpha,\beta=e,\mu,\tau$ and 
\begin{equation}
H=U\!\left(\begin{array}{ccc}0 & & \\ 
& \Delta_{12} & \\ 
& & \Delta_{13}\end{array}
\right)\!U^{\dagger}+
\sum_{n=1}^{\infty} a_n\left(\frac{E}{E_0}\right)^{2n-1}\!
\left(\!\begin{array}{c}\cos\zeta \cos\theta_L \\ 
\cos\zeta \sin\theta_L \\ 
\sin\zeta\end{array}\!\right)\!
\left(\begin{array}{ccc}\cos\zeta \cos\theta_L & 
\cos\zeta \sin\theta_L & \sin\zeta\end{array}\right).
\label{H}
\end{equation}
Here, $\Delta_{ij}\equiv\Delta m^2_{ij}/2E$, $i,j=1,2,3$ where $\Delta m^2_{ij}$ 
are the neutrino mass-squared differences and $U$ is
the leptonic mixing matrix, while $a_n\equiv b_n/E_0$ (the $a_n$
coefficients have dimensions of mass). The mixing angles $\zeta$ and
$\theta_L$ are defined by Eq.~(\ref{H}). They characterize the flavor
content of $\nu_L$, the neutrino that couples to the LIV sector. We
have assumed, for simplicity, that there are no CP-odd parameters in
the LIV sector. 

All information we need to analyze the neutrino data is included in
Eq.~(\ref{H}). Before proceeding, it is convenient to describe how we
hope to accommodate all the neutrino data:
\begin{itemize}
\item 
For large energies and short distances, $|\Delta_{ij}| L\ll 1$,
all oscillation behavior is governed by the LIV terms. These will be
fixed by the LSND anomaly plus other ``short-baseline'' constraints.
\item 
For ``solar'' oscillations, we will take advantage of the energy
dependence of the LIV term in order to make sure that these are
negligible at typical solar neutrino energies. In particular, we will
find that all $a_n$ will vanish for sufficiently small values of $n$.
\item 
For ``atmospheric'' oscillations, LIV effects will turn out to be very
small.    At large enough energies, the neutrino 
that couples to the LIV sector is a Hamiltonian eigenstate, while,
because the LIV part of the Hamiltonian has rank one (two
zero eigenvalues), the other two eigenvalues are still of order $\Delta_{12},\Delta_{13}$. 
We will choose $\zeta$ and $\theta_{L}$ so that the ``LIV neutrino''
is mostly $\nu_e$, and standard $\nu_\mu\leftrightarrow\nu_{\tau}$
oscillations governed by $\Delta_{13}$ remain virtually undisturbed for atmospheric-like 
neutrino energies. 
\end{itemize}

\section{Constraints from the Neutrino Data}
\label{sec:fit}

Here we explore choices for the LIV parameters that may allow one to
accommodate all neutrino data. We do not aim at performing a global
fit to all data, but concentrate on qualitatively understanding and satisfying 
the relevant constraints.

For LSND-like energies and baselines ($E\sim 50\;$MeV and $L\sim
30\;$m), the $\nu_{\mu}\to\nu_e$ transition probability is easy to
compute. At this point, we further assume that the energy
dependence is such that, at least for LSND-like energies,
\beq
\sum_{n=1}^{\infty} a_n\left(\frac{E}{E_0}\right)^{2n-1} \sim 
a_N\left(\frac{E}{E_0}\right)^{2N-1},
\eeq
where $a_N$ is the smallest nonzero $a_n$ term. The reason for
this will become clear when we look at the solar neutrino data. We
will comment on the behavior of the sum at large energies in due time.
It is easy to compute all $P_{\mu\alpha}$ in the limit
$|\Delta_{13}|L\ll 1$:
\begin{eqnarray}
P_{\mu e}&=&\sin^22\theta_L\cos^2\zeta\sin^2
\left(a_N\left(\frac{E}{E_0}\right)^{2N-1}\frac{L}{2}\right), \\
P_{\mu\tau}&=&\sin^22\zeta\sin^2\theta_L\sin^2
\left(a_N\left(\frac{E}{E_0}\right)^{2N-1}\frac{L}{2}\right), \\
P_{\mu\mu}&=&1-4\cos^2\zeta\sin^2\theta_L
\left(1-\cos^2\zeta\sin^2\theta_L\right)
\sin^2\left(a_N\left(\frac{E}{E_0}\right)^{2N-1}\frac{L}{2}\right).
\end{eqnarray}

In order to fit LSND and KARMEN-2 data, we must make sure that, at
LSND, the oscillatory effects do not average out. Roughly
speaking, we would like to choose $a_N$ so that $P_{e\mu}^{\rm
LSND}/P_{e\mu}^{\rm KARMEN}\sim (L^{\rm LSND}/L^{\rm KARMEN})^2$, as
is the case in successful mass-induced oscillation fits to LSND and
KARMEN data \cite{karmen+lsnd}. Furthermore, we will assume that at
experiments with larger energies and longer baselines, oscillation
effects average out. Here, we will be particularly concerned with
failed searches for $\nu_{\mu}\to\nu_e$ and $\nu_{\mu}\to\nu_{\tau}$
at NOMAD \cite{nomad_e,nomad_tau}, CHORUS \cite{chorus}, and NuTeV
\cite{nutev}. These constrain, at the 90\% confidence level,
\begin{equation}
\sin^22\theta_L\cos^2\zeta<1.1\times 10^{-3},
\quad{\rm and}\quad
\sin^22\zeta\sin^2\theta_L<3.3\times10^{-4},
\label{accelerator_constraints}
\end{equation}
while the LSND data require
$\sin^22\theta_{L}\cos^2\zeta\gtrsim10^{-3}$.  We therefore choose
$\sin^22\zeta=0$ to avoid the NOMAD constraint \cite{nomad_tau}, and
are forced to choose $\sin^22\theta_L=1.1\times 10^{-3}$ in order to
avoid the NuTeV constraint \cite{nutev} and to maintain hope that
there is a passable fit to the LSND data.

Solar neutrino data, combined with those from KamLAND, are very well
explained by solar matter-affected $\nu_{e}\to\nu_x$ oscillations
(where $\nu_x$ is some linear combination of $\nu_{\mu}$ and
$\nu_{\tau}$). This is clearly visible in the fact that, for $^8$B
neutrinos, $P_{ee}$ is significantly less than one half, which, given
our current understanding of neutrino masses and mixing, can only be
obtained if non-negligible solar matter effects are at work. In order
to avoid spoiling this picture, we require that, for solar-like
neutrino energies, LIV effects are negligible.

\begin{figure}[t]
\centerline{\epsfig{width=0.5\textwidth, file=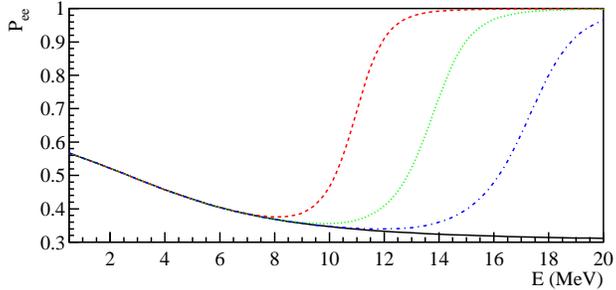}}
\caption{
Survival probability of solar neutrinos ($P_{ee}$) as a function of the solar
neutrino energy ($E$),  for $E_0=15$~MeV and different values of $a_N$. The
different curves correspond to $N=5$ and $a_5=0$~(solid, black line), $a_5=5\times 10^{-5}$~eV$^2$/MeV (dashed, red line), $a_5=5\times 10^{-6}$~eV$^2$/MeV (dotted, green line),
and $5\times 10^{-7}$~eV$^2$/MeV (dashed-dotted, blue line). See text for details.
\label{solar_a}}
\end{figure}

An order of magnitude estimate can be readily performed. The matter
potential in the sun's core is of order $A_{\rm sun}\sim 5\times
10^{-6}$~eV$^2$/MeV, so that significantly smaller $a_N$ values
guarantee that, for large $N$, $P_{ee}$ is undisturbed for energies
less than $E_0$. 
This can also be deduce from Fig.~\ref{solar_a} where
the survival probability of solar electron neutrinos as a function of
energy, for $E_0=15$~MeV and $N=5$, is shown. The different curves
correspond to $a_5$=0 (``normal'' LMA behavior), $a_5=5\times
10^{-5}$~eV$^2$/MeV, $a_5=5\times 10^{-6}$~eV$^2$/MeV, and
$a_5=5\times 10^{-7}$~eV$^2$/MeV. The relevant standard neutrino
parameters were set to $\Delta m^2_{12}=8\times 10^{-5}$~eV$^2$,
$\sin^2\theta_{12}=0.3$, and $\sin^2\theta_{13}=0$. In order to
compute $P_{ee}$, we use the formalism outlined in \cite{ether}, to
which we refer for all the relevant details.

Since the $\nu$SM successfully describes solar neutrinos up to the ``end'' of
the $^8$B neutrino spectrum (around 15~MeV), its predictions should not be 
dramatically modified for these neutrino energies. Then, from the figure, we see
that, for $E_0=15$~MeV, solar data require $a_5\lesssim
10^{-6}$~eV$^2$/MeV. This estimate is qualitatively independent of
$N$. For larger values of $N$, the same behavior is observed, except
that the transition between LMA-like behavior and $P_{ee}=1$ is more
abrupt. We conclude that if we choose $E_0 \sim 15$ MeV and $a_N\sim 10^{-6}$~eV$^2$/MeV, 
observable effects in solar data should be safely absent.  Once such
parameter choices are made, KamLAND data is completely oblivious to
LIV effects, given that relevant reactor neutrino energies are less
than 10~MeV.

Next we consider atmospheric neutrinos. For the values of $N$ in which
we are interested, LIV effects for $E\ge 100$ MeV are very large and
$\nu_L$ is basically decoupled from the other two
neutrinos. Thus, we are left with a standard effective two-flavor
oscillation scheme between these two states, governed by the
oscillation frequency $\Delta_{13}$. Since we choose $\nu_L$ to be
mostly $\nu_e$, the other two neutrinos are mostly linear combinations
of $\nu_\mu$ and $\nu_\tau$, such that $\nu_{\mu}\to\nu_{\tau}$
oscillations proceed as if there were no LIV effects. Effects due to
the deviation of $\nu_L$ from a pure $\nu_e$ state can be readily
computed, and turn out to be very small. We have checked that, for the
parameter values considered here, such deviations are within the
current experimental uncertainties.\footnote{These effects are indeed very small, and
it is not clear whether one will be able to probe them in future atmospheric
neutrino experiments.}

All considerations above constrain
\begin{equation} 
P_{\mu e}^{\rm LSND}\sim10^{-3}\sin^2
\left[2.54\times10^{-6}\left(\frac{E}{15~\rm MeV}\right)^{2N-1}
\left(\frac{L}{\rm m}\right)\right],
\end{equation}
so that $N$ is the only left-over parameter. Since we wish to avoid
averaged-out oscillations for typical LSND energies and baselines (if
at all possible), the best one can do is to choose the oscillation
phase to be close $\pi/2$ for typical LSND parameters, $L\sim 30$~m
and $E\sim 50\;$MeV. This translates into $2N-1=9$, or
$N=5$.\footnote{For smaller values of $N$, there is ``no hope'' of
reconciling solar and LSND data. A detailed discussion was presented
for $N=1/2$ (energy independent LIV effect) in
\cite{ether}.} This is far from providing a good fit to the LSND data,
for a couple of reasons. $P_{\mu e}^{\rm LSND,max}\lesssim 10^{-3}$,
which, even in the case of constant $P_{\mu e}^{\rm LSND}$, already
provides a mediocre explanation to the LSND anomaly. To add insult to
injury, because of the acute energy dependence, $P_{\mu e}^{\rm LSND}$
falls very rapidly with energy, so that, on average, $P_{\mu e}\ll
10^{-3}$.

Even if one overlooks the fact that we failed to accommodate all
neutrino data, there is another issue we need to deal with. For values
of the LIV parameters discussed above, and assuming that all other
$a_n$ to vanish, the neutrino dispersion relation starts to differ
significantly from that of a standard, ultra-relativistic particle for
$|\vec{p}|$ values larger than a few GeV. One way to avoid as
uncomfortable a situation is to postulate that, for large values of
$|\vec{p}|^2$, $f$ either stops increasing or decreases. We will now
argue that both possibilities not only avoid a dramatic modification
of the neutrino dispersion relation for GeV energies, but allow one to
contemplate accommodating all neutrino data more comfortably.

We will briefly investigate two examples. One is to assume that 
\begin{equation}
\sum_{n=1}^{\infty} a_n\left(\frac{E}{E_0}\right)^{2n-1}=
\kappa\left[\tanh\left(\frac{E}{15~\rm MeV}\right)\right]^{2l+1},
\label{tanh_eq}
\end{equation}
where $l$ is a positive integer, $\kappa$ is a constant coefficient
and we chose to express the energy in ``units'' of 15$\;$MeV, as before.
In this case, the LIV potential is at most equal to $\kappa$ (in the
limit $E\gg 15$~MeV), and we choose $\kappa$ so that, for LSND-like
neutrino energies and baselines, $\kappa L\sim 1$, which leads to
$\kappa\sim 0.02$~eV$^2$/MeV. In order to accommodate the solar data,
we need to choose $l$ so that $\kappa\tanh^{2l+1}(1)\lesssim
10^{-6}\;$eV$^2$/MeV. This translates into $l\gtrsim 16$. It remains to
discuss the faith of NuTeV (plus CHORUS and NOMAD) neutrinos. If we
fix $\zeta=0$, $P_{\mu\tau}$ vanishes and, since here $E\gg 15$~MeV,
\begin{equation}
P_{\mu e}\simeq \sin^22\theta_L\sin^2
\left[5.1\left(\frac{\kappa}{0.02~\rm eV^2/meV}\right)
\left(\frac{L}{100~\rm m}\right)\right],
\end{equation} 
{\it i.e.}, the oscillation length is energy independent.  In reality, 
we anticipate that  oscillatory effects average out because
different neutrinos propagate different distances between production
and detection. Given that decay tunnels and detectors are several tens
of meters long, and the oscillation lengths we are interested in are
of order 30~m, it sounds like a reasonable assumption. Under this
assumption the constraints listed in
Eq.~(\ref{accelerator_constraints}) apply.\footnote{If oscillation
effects did not average out, we would be allowed to choose $\kappa$ so
that $\nu_{\mu}$ to $\nu_e$ oscillation effects at these experiments
are minimized, hence loosening the constraints.}

\begin{figure}[t]
\centerline{\epsfig{width=0.5\textwidth, file=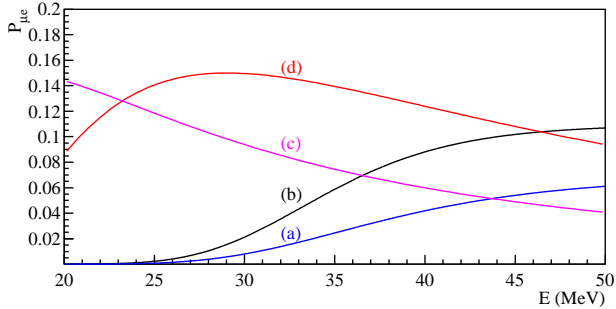}}
\caption{$\nu_{\mu}\to\nu_e$ transition probability ($P_{\mu e}$)
as a function of energy $(E)$ at (a) KARMEN ($L=18$~m) and (b) LSND
($L=30$~m), using the $\tanh$ ansatz (Eq.~\ref{tanh_eq}), compared
with those obtained with marginal mass-induces oscillation solutions
((c) at KARMEN and (d) at LSND). See text for details.
\label{tanh_fit}}
\end{figure}

Fig.~\ref{tanh_fit} depicts $P_{\mu e}$ as a function of energy for
(a) $L=18\;$m (typical of KARMEN baselines) and (b) $L=30\;$m (typical
of LSND baselines) for $\kappa=0.02\;$eV$^2$/MeV, $l=16$, and
$\sin^22\theta_L=1.1\times 10^{-3}$. Also depicted ((c)+(d)) are
equivalent two-flavor mass-induced oscillations for $\Delta
m^2=1.5$~eV$^2$ and $\sin^22\theta=1.5\times 10^{-3}$,
parameter choices that should fit the combined LSND+KARMEN data at the
99\% confidence level
\cite{karmen+lsnd}.\footnote{Recall, however,
that the smallest allowed value of $\sin^22\theta$ at a given
confidence level depends on the statistical analysis method employed
\cite{karmen+lsnd}.} This choice of parameters almost
provides a good fit to all data, were it not for the fact that
$\sin^22\theta_L$ is constrained to be a little too small.

A much more satisfactory fit could be obtained if the constraints
listed in Eq.~(\ref{accelerator_constraints}) did not apply. In that
case a good fit to the LSND data can already be obtained for
$\sin^22\theta_L\gtrsim 2\times 10^{-3}$.  It is plausible that a more
detailed analysis of NuTeV, NOMAD, and CHORUS data, beyond the
intentions of this short paper, would reveal that it is possible to
loosen Eq.~(\ref{accelerator_constraints}) by some factor of order
one, but understanding whether this is indeed the case is far from
trivial.  Among the issues that need to be considered carefully is the
fact that detectors and decay pipes are not negligibly small compared with
the typical baselines, and one needs to average over their dimensions
taking into account the exponential decay profile of the parent pions
and kaons that originate the different neutrino beams.

Finally, we mention the most radical change. Instead of considering LIV effects that behave  
like Eq.~(\ref{tanh_eq}), we postulate that these
are best represented by a function that is negligible at $E\lesssim
15$~MeV and $E\gtrsim 1$~GeV, while relevant for LSND energies
``in between.'' In this case, almost by construction, all neutrino
data can be comfortably accommodated. There is one concern that arises
from atmospheric data. If $|f|$ falls for energy values above 100~MeV,
at some energy one would encounter resonant $\nu_{\mu}$ to $\nu_e$
transitions. While the resonance is expected to be very narrow and
hence ``harmless,'' it can be avoided altogether if one chooses
$\zeta$ and $\theta_L$ so that $\nu_L$ is orthogonal to
$\nu_3$. 

Clearly, this possibility is even more tailor-made toward
resolving the LSND anomaly. Moreover, as we will briefly comment
later, we have no simple argument that justifies suppressing LIV
effects at large energies other than very nontrivial fine-tuning.

\section{Summary and Remarks}
\label{sec:end}

We have explored a new three-flavor fit to all neutrino data,
including those from the LSND experiment. We postulate that one of the
neutrinos couples to a source of Lorentz-invariance violation (LIV),
so that its dispersion relation is modified in such a way that
oscillation effects proportional to $\sin^2(E^nL)$ are present for
LSND neutrinos (see Eq.~(\ref{H})). For positive $n$ values, we see
that it is possible to accommodate solar and LSND data, while
atmospheric data is easily accommodated as long as the LSND mixing
angle and ``oscillation length'' are small.  Curiously enough, the
strongest constraints to this approach are provided by searches for
$\nu_{\mu}\to\nu_e$ oscillations at ``small'' $L/E$. Constraints
obtained by NOMAD, CHORUS and NuTeV
\cite{nomad_e,chorus,nutev} seem to allow only a mediocre fit to all
neutrino data, unless the LIV parameters are chosen so that LIV
oscillations are only relevant at LSND-like energies
(30$\;$MeV~$\lesssim E\lesssim$~100$\;$MeV).

The model explored here is, by construction, very finely-tuned.
Indeed, all manifestations of it are. The point where the LIV term
becomes important is immediately below LSND energies and above typical
solar neutrino energies. We were forced to choose very steep functions
of the neutrino energy for energies between few MeV to tens of MeV.
Clearly, this class of models is not motivated, but just serves as an
example of how nontrivial it is to accommodate the LSND anomaly when
the rest of the world neutrino data is included. In
order to obtain a ``natural'' fit, we are all but forced into
postulating that LIV effects are only present at LSND.

The character of the LSND anomaly will, hopefully, be clarified with
new experimental data.  The MiniBooNE experiment \cite{miniboone},
currently taking data at Fermilab, was constructed in order to test
whether the LSND anomaly is a manifestation of neutrino flavor
transitions. As far as the different scenarios discussed here are
concerned, MiniBooNE, which studies muon-type neutrinos with energies
of several hundreds of MeV, is expected see either no ``new physics''
effect -- if the LIV are indeed concentrated at LSND energies only --
or a rather small signal consistent with ``averaged out''
$\nu_{\mu}\to\nu_e$ oscillations. On the other hand, if MiniBooNE were
to see a significant, energy dependent signal for $\nu_e$ appearance,
LIV models of the type discussed here would be significantly
disfavored. Hence, in the case of no (significant) sign of new physics
at MiniBooNE, we would be unable to rule out the kind of LIV effects
discussed here.  Conclusive information capable of resolving the LSND anomaly
and addressing the models discussed here could be obtained, on the
other hand, in new experiments with lower energy neutrinos, such as
``long-baseline'' studies of muon or pion decay at rest (see, for
example, \cite{louis}).

We find it remarkable that so many distinct attempts to address the
LSND anomaly by augmenting the $\nu$SM fail because of distinct
aspects of the world neutrino data (see, for example, \cite{strumia},
for a nice overview). Some $L/E$ effects, for example, are
disfavored by atmospheric and solar data (as in the case of ``2+2''
spectra) or by combined failed searches for $\nu_{\mu}$ and $\nu_e$
disappearance at ``short'' baselines (as in the case of ``3+1''
spectra). $L$-independent solutions are disfavored by the KARMEN
experiment (see, for example, \cite{karmen_mu} for a specific
example), and we find that $L\times E^n$ effects are ultimately
disfavored by failed $\nu_{\mu}\to\nu_e$ searches at NOMAD, CHORUS,
and NuTeV (this was also mentioned in \cite{strumia}).

A few model-building comments are in order. We need to introduce a term
that violates Lorentz invariance and couples to neutrinos. Instead of
providing a full model, we simply make the following general remarks. 
Consider a scalar field $\phi$ whose time-derivative (in the preferred reference frame)
acquires a vacuum expectation value:
\begin{equation}
\langle \partial_\mu \phi \rangle = \delta_{0\mu} M,
\end{equation}
and we further impose a $Z_{2n}$ symmetry on $\phi$. This symmetry may
be softly broken in the $\phi$ potential, but is assumed to be
(almost) exact when considering the couplings to neutrinos, so that
only terms that scale like $E^{2n}$ (and its powers) are
allowed. Hence, symmetry arguments can be used to explain why only
large $N$ values contribute.  A similar argument applies for LIV
effects proportional to $[\tanh(E)]^m$ as far as the exponent $m$ is
concerned, but we cannot use symmetry arguments to ``explain'' why
infinite series of higher dimensional LIV operators would combine into
either a $\tanh(E)$ form, or why effects should be enhanced at
neutrino energies close to 50~MeV (but severely suppressed at energies
above or below 50~MeV).

In summary, we succeed in identifying a three-neutrino, LIV solution
capable of accommodating all neutrino data. The price we had to pay in
order to do that was quite high: not only were we forced to impose
that the LIV effects are very strongly energy-dependent for energies
around 20~MeV, we were also forced to postulate that LIV effects either
``flatten out'' for energies larger than 100~MeV (in which case the
fit to LSND data is mediocre) or peak at LSND energies and quickly
``disappear'' for $E\gtrsim 1\;$GeV.

We do not advocate our model as a solution to the LSND anomaly, but
rather as another evidence for the ``real'' LSND puzzle: how can the
effect observed at LSND be due to physics beyond the $\nu$SM if all
other experiments (``neutrino'' or otherwise) have failed to observe
related ``new physics'' effects? If the integrated literature on the
subject is to be taken as a good indicator, one would, naively,
conclude that the answer to the question above is `it cannot.' It
seems unfair, however, to conclude that a new physics interpretation
of the LSND anomaly does not exist simply because there seems to be no
elegant solution to the LSND anomaly. It may well be we just have not
found one yet.

\section*{Acknowledgments}
We thank Andy Cohen, Shelly Glashow, and Concha Gonzalez-Garcia for
helpful discussions, and Yossi Nir for comments on the manuscript.
The work of AdG is sponsored in part by the US Department of Energy
Contract DE-FG02-91ER40684. The work of YG is supported in part by the
Israel Science Foundation under Grant No.  378/05.

 \end{document}